\documentstyle[aps,twocolumn,prl,epsf]{revtex}

\newcommand{\beqar}{\begin{eqnarray}}
\newcommand{\eeqar}{\end{eqnarray}}
\newcommand{\beq}{\begin{equation}}
\newcommand{\eeq}{\end{equation}}

\begin{document}
\twocolumn[\hsize\textwidth\columnwidth\hsize\csname
@twocolumnfalse\endcsname

\title{ Generalized Phase Field Model for Computer Simulation of
Grain Growth in Anisotropic Systems}
\author{A. Kazaryan$^1$, Y. Wang$^2$, S.A. Dregia$^2$ and Bruce R. Patton$^1$}
\address{$^1$Department of Physics,
         $^2$Department of Materials Science and Engineering,
	 The Ohio State University,
         Columbus, OH 43210}

\date{\today}

\maketitle

\begin{abstract}
We study the dynamics and morphology of grain growth with anisotropic
energy and mobility of grain boundaries using a generalized phase
field model. In contrast to previous studies, both inclination and
misorientation of the boundaries are considered. The model is first
validated against exact analytical solutions for the classical problem of
an island grain embedded in an infinite matrix. It is found that grain
boundary energy anisotropy has a much stronger effect on grain shape
than that of mobility anisotropy. In a polycrystalline system with
mobility  anisotropy, we find that the system evolves in a
non-self-similar manner and grain shape anisotropy develops. However,
the average area of the grains grows linearly with time, as in an
isotropic system. 
\end{abstract}

\draft 
\vskip1.5pc]

\par

The effect of anisotropy in energy and mobility of grain boundaries on
the kinetics of grain growth and morphological evolution is a
relatively unexplored problem. Studies have shown that energy and
mobility of a grain boundary depend on the misorientation
between the two crystals and the inclination of the grain
boundary \cite{gb_conf}. In addition, phenomena  as segregation of
impurities \cite{defects} or presence of a liquid phase \cite{liqid} at the
grain boundaries may also result in anisotropy of both energy
and mobility.

\par
Although most computer simulations of grain growth have
been performed for isotropic cases
\cite{srolovitz,cocks,atkins,is_chen_1,is_fan}, anisotropy in
grain boundary properties has been introduced in a number of simulations,
mainly by the Monte Carlo method 
\cite{novikov,humphreus,mehnert,grest,ono,yang-chen,cai-welch}.
For example in the study of texture development during grain growth
\cite{novikov,humphreus,mehnert}, grains are divided into two
types and the contacts between them form three kinds of grain
boundaries of either
small or large misorientations. To take into account the full range of
grain orientations, more general approaches have been proposed
\cite{grest,ono,yang-chen,cai-welch}. However, all these models
consider either misorientation or inclination dependence of grain
boundary properties.  A simple dislocation model of
the grain boundary shows that both energy and mobility of the boundary
could depend strongly on both misorientation and  inclination
\cite{read_book}. Furthermore, in the models that deal with inclination
dependence \cite{yang-chen,cai-welch} only a few inclinations
are considered and the use of only first neighbors for the calculation
of boundary inclination results in an intrinsic energy and mobility
anisotropy associated with the discrete lattice used in the
simulations \cite{hrs}. 

\par

In this paper, to study the kinetics and morphology of grain growth in
anisotropic systems, we extend the phase field approach to take into
account both inclination and misorientation dependence of grain boundary
energy and mobility. The phase field method has been successfully applied
for computer simulation of isotropic grain growth\cite{is_chen_1,is_fan},
phase transformations \cite{yw-chen}, and solidification \cite{karma}. In
this model, the polycrystalline microstructure is described by a set of
non-conserved order parameter fields $(\eta_1, \eta_2, \ldots, \eta_p)$,
each representing grains of a given crystallographic orientation.
Microstructural evolution of a polycrystalline system is characterized by
the spatio-temporal evolution of the order
parameters, via the Ginzburg-Landau-type kinetic equations:
\beq
\frac{\partial\eta_{i}}{\partial t}=-L\frac{\delta
F}{\delta\eta_{i}} \hspace{0.2cm}, \hspace{0.6cm} i=1, 2,\ldots, p
\label{eq:Gins_Land}
\eeq
where $L$ is the kinetic coefficient characterizing the grain boundary
mobility, and $F$ is the free energy functional of the form:
\beqar
\nonumber F=F_{0}+\int d^{3}r [f(\eta_1( \textbf{r}), \eta_2(
\textbf{r}),\ldots, \eta_p( \textbf{r})) \\
+ \frac{k}{2} \sum_{i=1}^{p} (\nabla\eta_i)^2]
\label{eq:free_en}
\eeqar
where $f(\eta_i)$ is the local free energy density and $k$ is the
gradient coefficient, which together determine the width and energy of
the grain boundary regions.

\par

Several attempts have been made to introduce anisotropic boundary
properties into the  phase field formulation. For example, in the phase
field models of antiphase domain growth and solidification,
anisotropy has been
introduced by using multiple order parameters based on the underlying
crystal symmetry \cite{braun,agk}. However, application of these models to
grain growth is very difficult due to the fact that the
description of the grain  boundary is much more complicated, and a
quantitative description of a general grain boundary is still lacking.
Recently, Kobayashi \textit{et al.} \cite{kwc} and independently Luck
\cite{luck} suggested a different way to extend the phase field model to
include anisotropy in grain boundary properties. In their approach,
anisotropy is incorporated by introducing an additional variable that
describes spatial orientation of the grains. So far, the model has
been applied to only quasi-one-dimensional systems with misorientation
dependence of grain boundary energy.

\begin{figure}[t]
\epsfysize=3.7cm
\centerline{\epsffile{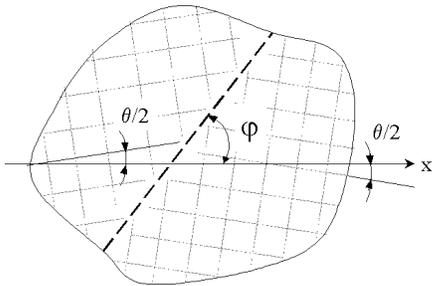}}
\vspace{0.3cm}
\caption{Schematic drawing of a tilt grain boundary formed by two crystals
with misorientation $\theta$ and inclination $\phi$, measured from the
symmetric tilt boundary}
\label{Fig:pic_r_s}
\end{figure}

In contrast, a simple phenomenological approach has been successfully
used to describe the surface 
energy and mobility anisotropy of crystal surfaces in phase field modeling
of crystal growth, where the gradient coefficient $k$
and the kinetic coefficient $L$ have been formulated as functions of
crystal surface orientations \cite{karma,nist_93}. In the current
paper a similar approach is 
used to describe grain boundary energy and mobility anisotropy. For the
sake of simplicity, we consider tilt grain boundaries between
two-dimensional crystals with square lattices, which
(ignoring rigid-body translations) can be described by two parameters:
relative orientation $\theta$ of the grains (misorientation) and spatial
orientation $\phi$ of the boundary with respect to the reference coordinate
system (inclination). For crystals with four-fold symmetry and based
on simple dislocation models \cite{read_book}, the energy of a low-angle
tilt boundary ($\theta \leq 20^{0}$) can be approximated as: 
\beq
E(\theta,\phi) = E_{0} ( \left| cos(\phi) \right| + \left| sin(\phi)
\right| ) \theta (1 - ln(\theta / \theta_{m}))
\label{eq:r_s_eq}
\eeq
where $\phi=0$ corresponds to the symmetric tilt boundary
(see Fig.\ref{Fig:pic_r_s}), $\theta_{m}$ is the misorientation at
which energy is maximum
and $E_0$ is a constant. The grain boundary energy can be
related to the gradient coefficient by $k \propto E^{2}(\theta, \phi)$
\cite{cahn_79}. However, as noted by McFadden \textit{et al.}, the
gradient coefficient must be a
differentiable function with respect to inclination for the
Ginzburg-Landau equations to be properly defined \cite{nist_93}. Thus in
our simulations the grain boundary energy has been taken in the form:
\beq
E(\theta,\phi) = E_{0} (1 - \delta_E cos(4 \phi) ) \theta (1 - ln(\theta
/ \theta_{m} ))
\label{eq:our_en}
\eeq
which maintains four-fold symmetry in $\phi$, where $\delta_E$ is a
phenomenological parameter serving as a measure of the degree of 
anisotropy. The inclination dependence of the energy is similar to the
one used in the phase field model of solidification \cite{karma}. By
choosing $\delta_E = 0.24$ the same anisotropy ratio as
in Eq. (\ref{eq:r_s_eq}) is obtained, where the anisotropy ratio $r$
is defined as the ratio of the largest to the smallest grain boundary
energy at a fixed misorientation $\theta_0$: $r = max{E( \phi,
\theta_0 )} / min{E(\phi, \theta_0)}$.  
\par

In the phase field approach to grain growth, grain boundary mobility is
characterized by the kinetic coefficient $L$. In contrast to the grain
boundary energy, values of the grain boundary mobility are harder to
estimate. Qualitatively, in pure materials the higher the defect
concentration at a grain boundary the higher its energy as well as its
mobility. To investigate qualitatively the effect of mobility
anisotropy on the behavior of grain growth, we assume mobility to have
the same dependence on misorientation and inclination as the grain
boundary energy (Eq.(\ref{eq:our_en})), with a different
phenomenological parameter $\delta_L$. By varying $\delta_E$ and
$\delta_L$ we can control the anisotropy ratio and investigate the
interplay between energy and mobility anisotropy.

\par

The free energy density in Eq.(\ref{eq:free_en}) has been taken in
the form \cite{is_chen_1,is_fan}:
\beq
f = \sum_{i=1}^{p} [-\frac{a_{1}}{2}\eta_{i}^2 +
\frac{a_{2}}{4}\eta_{i}^4 ] + \frac{a_{3}}{2} \sum_{i=1}^{p} \sum_{j >
i} \eta_{i}^{2} \eta_{j}^2 \hspace{0.2cm} .
\label{eq:free_en_density}
\eeq
In our simulations grain boundary misorientation was considered to be in the
range of $0 < \theta \leq \Theta$, where $\Theta < \theta_m$ Then,
misorientation $\theta_{ij}$ between grain $i$ and grain $j$
($i,j$ = $1 \ldots p$, where $p$ is the total number of order parameters
used in the simulation) is calculated as $|i - j| \cdot \Theta / (p-1)$.
The following values for the phenomenological parameters in
Eq.(\ref{eq:free_en_density}) have been used in the simulations:
$a_1 = 1.0$, $a_2 = 1.0$, $a_3 = 2.0$. Eq.(\ref{eq:Gins_Land})
was discretized using the second order Euler technique on a unit
square lattice with $dt = 0.1$.

\par

Below we present several applications of our model to systems
containing an island grain in an infinite matrix as well as to a
polycrystalline aggregate. 

\begin{figure}[t]
$\begin{array}{ccc}
\epsfysize=3.0cm \epsfbox{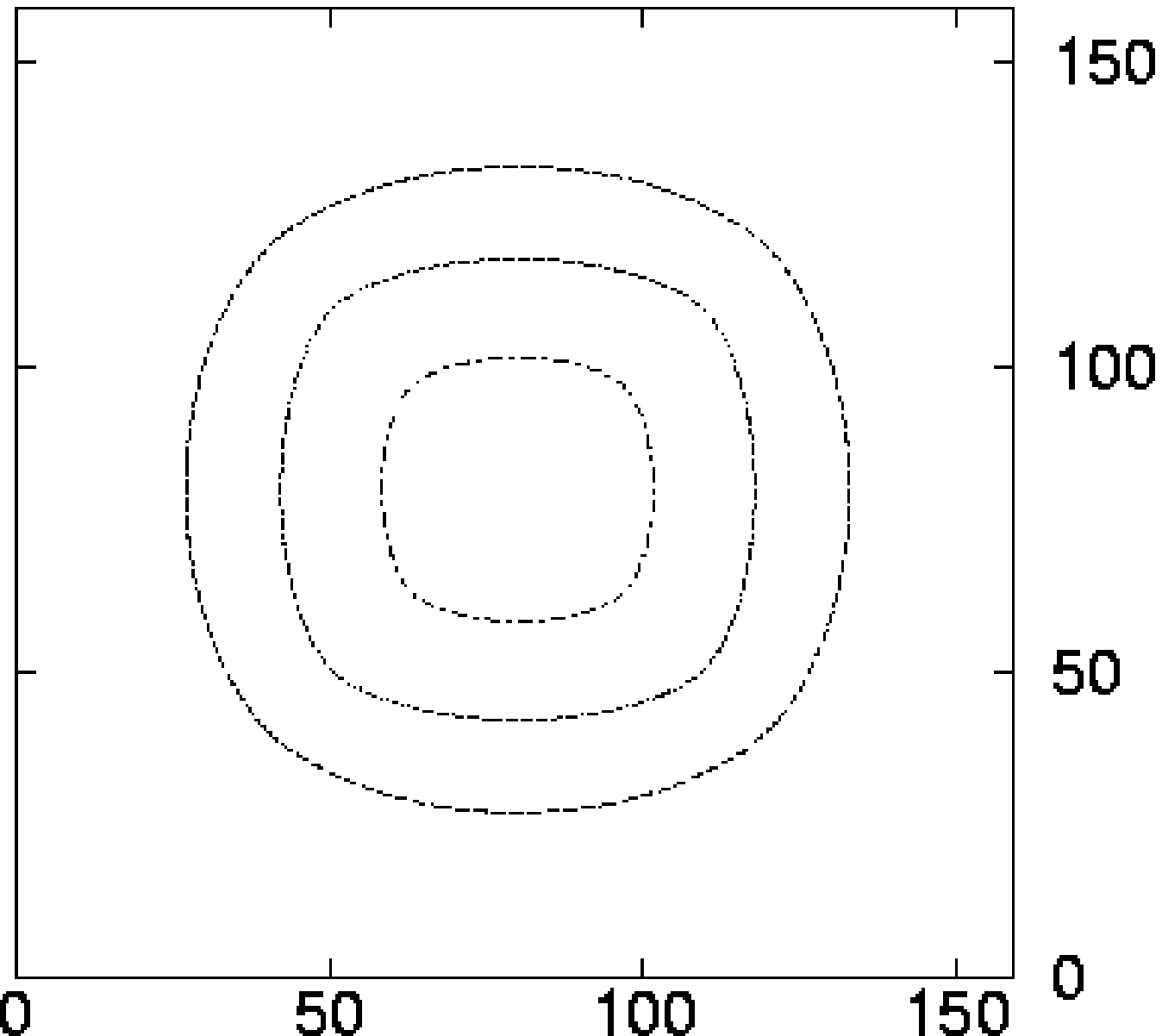} &
\epsfysize=3.0cm \epsfbox{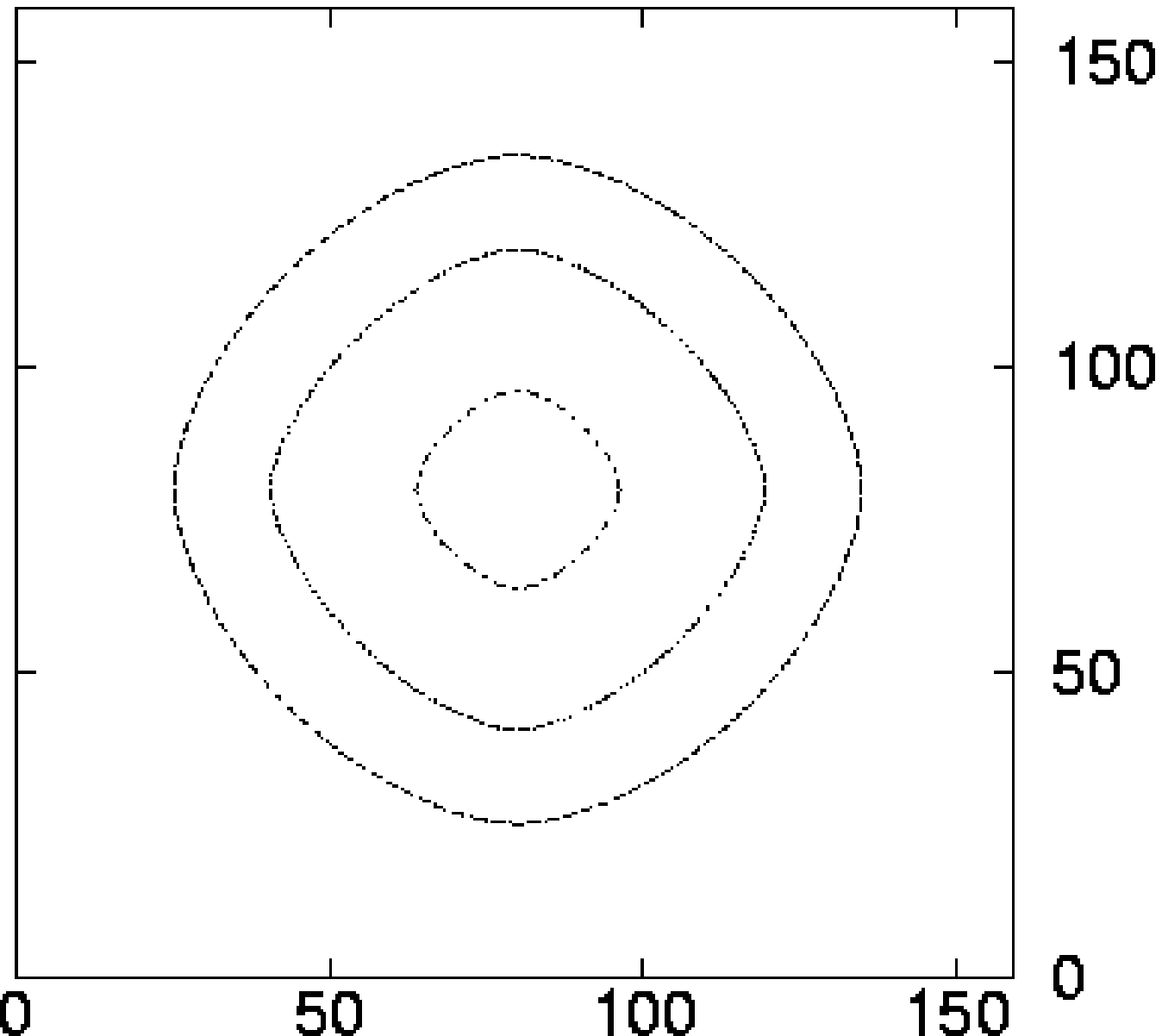} \\
(a) & (b) \\
\end{array}$
\begin{center}
$\begin{array}{c}
\epsfysize=3.0cm \epsfbox{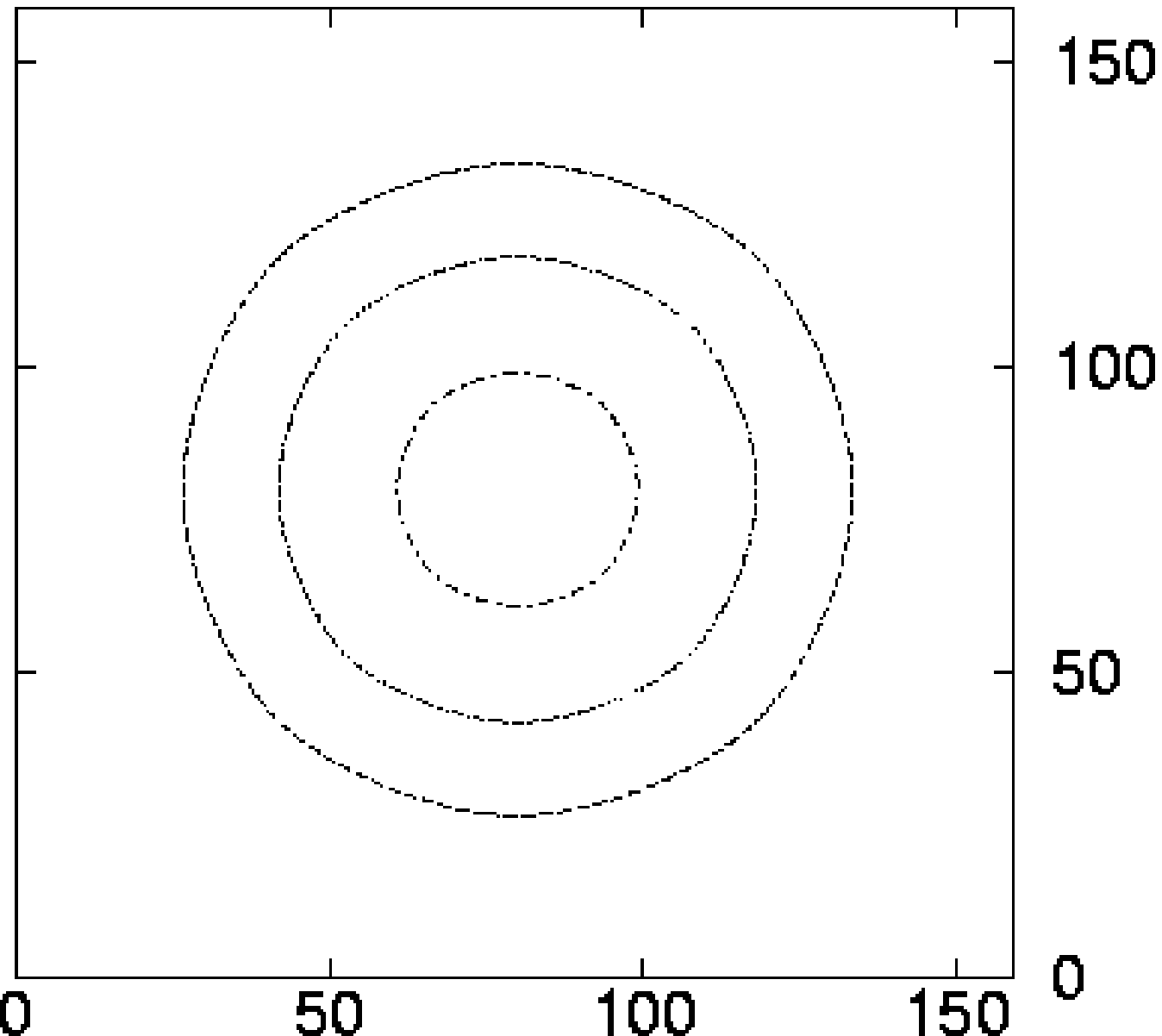} \\
(c) \\
\end{array}$
\end{center}
\caption{Shrinkage of a single island grain with (a) energy
anisotropy; (b) mobility anisotropy; (c) both energy and mobility anisotropy}
\label{Fig:grain_pic}
\end{figure}

\textit{Island grain with energy anisotropy.}
To validate the model against exact analytical results, we first
examine shrinkage of an island grain embedded in an infinite matrix.
If the grain shrinks in a ``self-similar'' manner, i.e. maintains its
shape while shrinking, Taylor and Cahn \cite{taylor_cahn} have shown
that the Ginzburg-Landau equations could be rewritten in the same form as
in the isotropic case. As a result, shrinkage kinetics  of a grain with
arbitrary but self-similar shape
should be the same as in the isotropic
case. Moreover, they have argued that depending on the form of the
gradient coefficient certain complications could occur in the solution
of the Ginzburg-Landau equations due to non-convexity in the Wulff
plots. Since we are mainly interested in the qualitative behavior of
the system and to avoid those complications, we have chosen a grain boundary
energy in the form (\ref{eq:our_en}) with $\delta_E =0.05$ and
$E_{0}=1/(1+\delta_E)$, which gives anisotropy ratio
$r_E=1.1$. The result of our simulations 
is presented in Fig.\ref{Fig:grain_pic}a. As expected, an initially
circular grain transforms to its Wulff shape and
then shrinks in a self-similar manner. The extracted kinetics indeed
shows linear decline of area with time, in agreement with the
theoretical predictions of Taylor and Cahn \cite{taylor_cahn}.

\par

\begin{figure}[t]
\epsfysize=3.7cm
\centerline{\epsffile{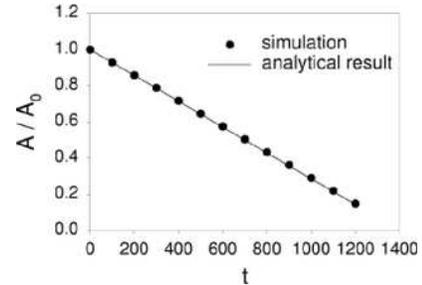}}
\vspace{0.1cm}
\caption{Shrinkage kinetics of an island grain embedded in an
infinite matrix: comparison between analytical solution
Eq.(\ref{eq:A_mobil}) (solid curve) and simulation (solid circles) in
the case of anisotropic grain boundary mobility. $A_0$ is the initial
area of the grain.}
\label{Fig:compare_mobil}
\end{figure}

\textit{Island grain with mobility anisotropy.}
Following the
theoretical analysis of Allen and Cahn \cite{cahn_79} for the
isotropic case, we obtain a similar relationship for the shrinkage
rate of a single island grain with anisotropic boundary mobility:
\beq
A(t) = A(0) - 2  \pi  k  \langle L \rangle_{\phi} t
\label{eq:A_mobil}
\eeq
where $A$ is the area of the grain, and $\langle \ldots \rangle_{\phi}$
represents an average over all possible inclinations. This expression is
valid for arbitrary particle shapes and the system does not
need to be in
a self-similar regime. In the simulations mobility has been
taken in the form (\ref{eq:our_en}) with $L_0 = 1/(1+\delta_L)$. In
contrast to the case of energy anisotropy, a small mobility anisotropy
($\delta_L$=$0.05$, which gives an anisotropy ratio $L_{max}/L_{min}=1.1$)
has little effect on the shapes of the grain. Indeed, to reach a
similar degree of anisotropy in shape(Fig.\ref{Fig:grain_pic}b), we
have to introduce a significantly higher mobility anisotropy with
$\delta_L$=$0.7$, which produces a relative mobility
difference (($L_{max}-L_{min})/L_{min}$) $\sim 500$\% (while only
$10$\% difference in energy is required). Experimental observations
have shown that the mobility difference could be orders of magnitude
larger than the energy diffrence for the grain boundaries
\cite{gb_conf,novikov,humphreus}. Finally, excellent agreement  
between the simulation results and 
the exact analytical solution (\ref{eq:A_mobil}) for the grain shrinkage
rate has been observed (Fig.\ref{Fig:compare_mobil}), which indicates
that discretization of the equations does not influence the kinetics.

\par

\textit{Island grain with both energy and mobility anisotropy.}
Next, we discuss the effect of the interplay between the energy and mobility
anisotropy of the grain boundary on the shape of a single
shrinking island grain. It should be noted that since the functions
$E(\phi)$ and $L(\phi)$ have been taken to have the same $\phi$
dependence, for a shrinking island 
grain energy and mobility anisotropy have opposite effects on the shape.
Boundaries with normals in $\frac{\pi}{2}n$ directions have the lowest
energies as well as lowest mobilities. Therefore, energy minimization would
prefer a shape bounded by the low energy boundaries
(Fig.\ref{Fig:grain_pic}a), while grain shrinkage
with anisotropic boundary mobility would produce a shape
bounded by the fastest moving boundaries (Fig.\ref{Fig:grain_pic}b).
As we mentioned earlier, a small anisotropy in the mobility
coefficient ($r_L = 1.1$) does not have much effect on the shape of
the grain; thus the grain shape is
dominated by energy anisotropy, giving a shape similar to
Fig.\ref{Fig:grain_pic}a. On the other extreme a very large
mobility anisotropy ($r_L = 40$) results in diamond-like shapes
(Fig.\ref{Fig:grain_pic}b).
In the intermediate range ($r_E$=$1.1$, $r_L$=$5.7$), a grain shape
with quasi eight-fold symmetry has been observed
(Fig.\ref{Fig:grain_pic}c), although  both energy and mobility as a
function of inclination have four-fold symmetry.

\par

\begin{figure}[t]
$\begin{array}{cc}
\epsfysize=3.5cm \epsfbox{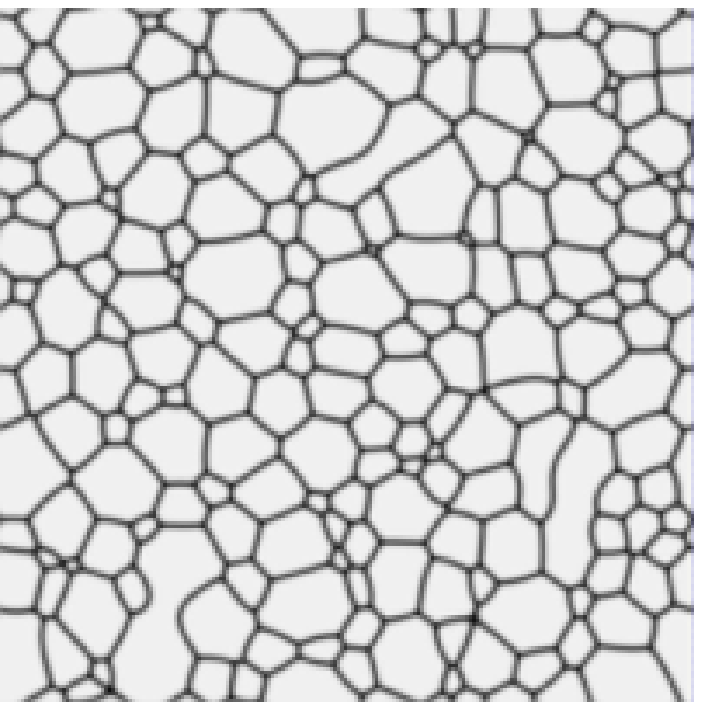} &
\epsfysize=3.5cm \epsfbox{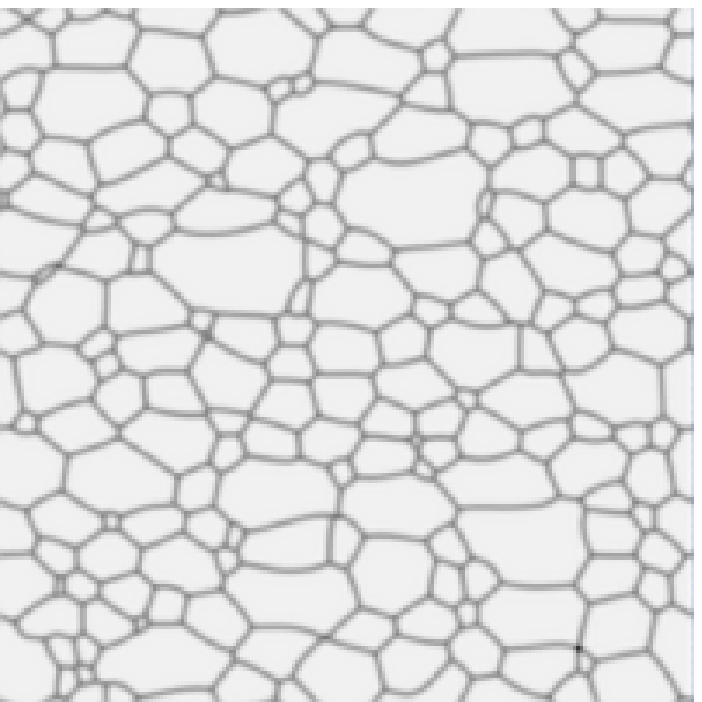} \\
(a) & (b)
\end{array}$
\caption{Typical microstructures obtained from computer simulations
with isotropic grain boundary energy and (a) isotropic and (b) anisotropic
boundary mobility with two-fold symmetry with respect to $\phi$}
\label{Fig:policrystall}
\end{figure}

\begin{figure}[t]
\epsfysize=3.7cm
\centerline{\epsffile{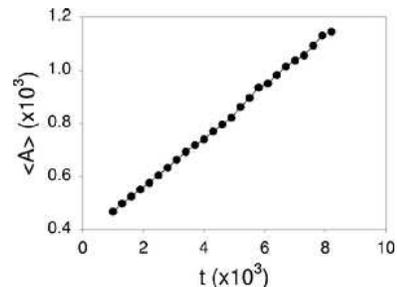}}
\vspace{0.1cm}
\caption{Time dependence of the average grain area in a
polycrystalline system with grain boundary mobility anisotropy.}
\label{Fig:area}
\end{figure}

\textit{Polycrystalline aggregate with mobility anisotopy.}
For a polycrystalline aggregate, Mullins has shown that the average grain
area grows linearly with time if the system evolves in a statistically
self-similar (scaling) manner (i.e. all configurations have identical
statistics when transformed to the same linear scale by uniform
magnification) \cite{mullins2}. However, self-similarity may not hold
when boundary 
properties (energy and/or mobility) are anisotropic \cite{mullins3}. Indeed,
our simulations performed on a polycrystalline system with grain boundary
mobility anisotropy have shown that grain shapes evolve in a
non-self-similar manner, e.g., shape anisotropy develops
(Fig.\ref{Fig:policrystall}b). The simulations were performed on $512 \times
512$ square lattice with $36$ order parameters. The simulations were started
from an isotropic polycrystalline microstructure consisting of $\sim
1000$ grains, obtained from nucleation and growth of crystals from a 
liquid phase in an isotropic system. Grain boundary mobility
anisotropy is introduced
according to $L=\theta  (1-ln(\theta / \theta_m))(1-\delta\cos(2
\phi))$, with $\theta_m=10^0$, $\Theta=5^0$ and $\delta_L=0.9$, which
has two-fold symmetry with respect to $\phi$ and $\phi=\pi/2$ is the
fastest growth direction. For 
comparison, the result obtained from the same initial microstructure in an
isotropic system is shown in Fig.\ref{Fig:policrystall}a. Comparing these
results it is clear that in the anisotropic case grain shape
anisotropy develops. Quantitative analysis of
Fig.\ref{Fig:policrystall}b has shown that the inclination
distribution of grain boundaries (i.e., boundary length as a function
of inclination) is no longer uniform, which
suggests that the microstructure is no longer self-similar. However, the
average grain area still grows linearly with time  as shown in
Fig.\ref{Fig:area}. It can be shown analytically \cite{ak_yw} that
even though  self-similarity is not satisfied when grain boundary mobility is
anisotropic, the average grain area may still grow linearly with time.

\par

In conclusion, we have studied the dynamics and morphology of grain growth
with both energy and mobility anisotropy of grain
boundaries using a generalized phase field model that incorporates both
inclination and misorientation dependence of grain boundary properties. For
a single island grain embedded in an infinite matrix we found
that a small anisotropy in grain boundary energy can produce strongly
anisotropic grain shapes. In contrast, the mobility anisotropy needs to be
significantly stronger to produce similar morphologies. For a
polycrystalline aggregate with mobility anisotropy, we found that the
average area of the grains grows linearly with time, as in the
isotropic case, even though grain shape anisotropy develops, which indicates
that the microstructural  evolution is no longer self-similar.

\par

We gratefully acknowledge the financial support of NSF under grant
DMR-9703044 (AK and YW) and NSF Center for Industrial Sensors and
Measurements under grant EEC-9523358 (AK and BRP).



\end{document}